# Protocol-Preserving Context Trimming for Agentic Workflows: Benefits, Failure Regimes, and Budget Guardrails

**HARISH GAGGAR**
Senior Staff Software Engineer, Intuit Credit Karma, Oakland, California, USA
Corresponding author: Harish Gaggar
Email: harish_gaggar@intuit.com

**Abstract**

Agentic large language model (LLM) systems increasingly rely on long interaction histories to preserve instructions, tool states, intermediate decisions, and unresolved dependencies, but unrestricted context growth raises computational cost and may reduce operational efficiency. This study evaluated protocol-preserving context trimming as a reliability-constrained approach for managing context in multi-step agentic workflows. Five trimming strategies; recency-based, relevance-based, summarization, protocol-aware trimming, and adaptive budget guardrails were assessed across multiple retained-context levels and workflow-complexity classes using task success, protocol adherence, valid tool calls, token savings, latency reduction, cascading failures, and critical context thresholds. Conventional strategies achieved approximately 60% mean token savings but showed lower task success (66.6-77.3%) and protocol adherence (85.5-88.6%). Protocol-aware trimming improved task success to 92.2%, while adaptive guardrails achieved 96.0% task success, 96.3% protocol adherence, and only 1.0% cascading failure with 56.0% mean token savings. Aggressive trimming at retained-context budgets of ≤25% increased failure odds 10.92-fold relative to budgets ≥50% ($p < 0.001$). Protocol-aware trimming produced 5.24-fold greater odds of successful completion than conventional approaches under aggressive budgets, while adaptive guardrails further increased success odds by 2.11-fold relative to fixed protocol-aware trimming ($p < 0.001$). Critical context thresholds also increased with workflow complexity, confirming that safe compression limits are task dependent. Overall, the findings show that reliable context reduction depends more strongly on preserving protocol-critical state than on maximizing token removal. Protocol-aware trimming combined with adaptive budget guardrails provides a promising framework for improving the efficiency, scalability, and reliability of long-horizon agentic systems.

## Introduction

The transition from conversational large language models (LLMs) to agentic systems has fundamentally altered the functional role of context [1]. In conventional question–answering systems, context primarily serves as an informational substrate for response generation. In agentic workflows, by contrast, context operates as a composite representation of memory, execution state, task constraints, tool interfaces, prior decisions, retrieved evidence, resource identifiers, and unresolved dependencies [2]. As workflows extend across multiple steps, tools, and external environments, this accumulated context increasingly resembles a dynamic computational state rather than a conventional dialogue history. Consequently, context growth is not merely a problem of redundant text or token inefficiency; it is a systems-level problem concerning how much operational state must remain available for an agent to continue a task correctly [3].

Expanding context windows partially relax capacity limitations, but they do not resolve the underlying trade-off between informational completeness and computational efficiency. Longer contexts increase token consumption, inference cost, and latency, while potentially reducing the relative salience of critical instructions embedded within extensive histories [4]. Prompt-compression approaches such as LLMLingua and LongLLMLingua have demonstrated that substantial reductions in input length can often be achieved without equivalent degradation in downstream performance. Likewise, hierarchical memory architectures such as MemGPT illustrate the potential of moving information between active and external memory rather than retaining complete interaction histories within the model context [5]. However, these approaches expose a more difficult question for agentic systems: whether information that appears redundant or semantically compressible can be removed without changing the executable meaning of the workflow.

This distinction between semantic preservation and operational preservation is central to reliable context management [6]. Conventional compression is typically judged by whether the reduced representation retains the principal meaning of the original text or supports comparable task performance. Such criteria may be insufficient for agents that operate tools, manipulate persistent state, or execute multi-stage plans. An agent may retain the general intent of a task while losing an exact resource identifier, exception, negative constraint, temporal dependency, authorization condition, or record of an action already performed [7]. These omissions can produce responses that remain linguistically coherent but are operationally invalid. The problem is compounded by the heterogeneous importance of contextual elements. A descriptive explanation from an earlier interaction may be safely summarized, whereas a file identifier, database key, schema constraint, pending commitment, or instruction not to modify an original resource may be effectively non-compressible [8]. Context value therefore cannot be inferred solely from length, recency, linguistic salience, or semantic similarity.

This study frames the resulting requirement as protocol preservation. A workflow protocol comprises the invariants that must remain intact for subsequent actions to remain valid, including instruction precedence, tool contracts, argument structures, entity bindings, permissions, state transitions, temporal ordering, resource identities, and user-defined constraints [9]. Protocol-preserving context trimming therefore differs from generic prompt shortening because its objective is not simply to maximize token reduction, but to reduce contextual load without altering the set of valid future actions available to the agent. This distinction becomes particularly important when agents interact with external environments, where errors may have persistent consequences. A conversational misunderstanding can often be corrected in a subsequent turn, whereas an incorrect API call, duplicated transaction, destructive file operation, or invalid state transition can change the environment and propagate errors through all later stages of the workflow [10].

The relationship between context reduction and workflow performance is therefore unlikely to be linear. Moderate trimming may remove duplicated explanations, obsolete observations, and low-value intermediate content, potentially improving both efficiency and instruction salience [11]. Beyond a critical threshold, however, additional compression may produce disproportionate reliability losses. Potential failure regimes include loss of instruction scope, deletion of unresolved constraints, state aliasing between similar entities, distortion of numerical values, loss of execution chronology, disappearance of negative requirements, premature summarization of evidence, and replacement of exact structured state with approximate natural-language representations [12]. These failures are especially concerning because they may remain undetected. An agent can continue producing apparently coherent actions even after the contextual information required for valid execution has been removed [13]. As a result, final-answer similarity or task completion alone may conceal important protocol violations.

A rigorous evaluation of context trimming must therefore extend beyond token savings. It should distinguish semantic performance from protocol adherence, state consistency, tool-call validity, constraint preservation, recoverability, and downstream error propagation [14]. This also challenges the widespread use of fixed compression ratios or generic context thresholds. The safe context budget is inherently task- and risk-dependent. Descriptive history may tolerate aggressive compression, whereas state-changing operations, permissions, identifiers, tool specifications, and unresolved dependencies may require near-lossless retention [15]. Context management is thus better understood as a constrained allocation problem in which available tokens are distributed according to operational criticality rather than uniformly across the interaction history [16].

Budget guardrails provide a practical mechanism for implementing this principle. Such guardrails can protect non-compressible protocol elements, reserve minimum context for active state, prevent trimming across unresolved dependencies, trigger retrieval or rehydration of archived information, and reduce compression aggressiveness during high-consequence stages [17]. The optimization objective consequently shifts from maximizing compression to minimizing context cost subject to an acceptable bound on protocol failure. This reframing is critical because extreme token efficiency has limited value if it undermines state integrity or produces irrecoverable operational errors.

Despite rapid progress in long-context inference, prompt compression, external memory, and autonomous agent architectures, these areas are still frequently evaluated as separate optimization problems. Compression research has primarily emphasized token reduction and semantic task performance, whereas evaluations of agentic systems have focused more strongly on reasoning, planning, and tool-use success. As a result, the point at which context reduction shifts from beneficial compression to protocol-destructive information loss remains insufficiently understood. This gap is particularly important for long-horizon workflows, where the consequences of trimming may emerge several steps after the relevant information has been removed. The present study therefore evaluates protocol-preserving context trimming as a reliability-constrained optimization problem. It is hypothesized that controlled removal of redundant and operationally inactive context will reduce token consumption and computational overhead without significantly affecting workflow completion or protocol adherence (H1). However, as the retained context budget falls below a workflow-specific critical threshold, the probability of protocol violations, state inconsistencies, invalid tool calls, and downstream execution errors is expected to increase disproportionately rather than linearly (H2). It is further hypothesized that protocol-aware trimming strategies that explicitly protect operationally critical information, including identifiers, unresolved constraints, tool specifications, permissions, state transitions, and temporal dependencies, will maintain significantly higher workflow reliability than relevance-, recency-, or generic summarization-based trimming at comparable token budgets (H3). Finally, the study hypothesizes that adaptive budget guardrails, in which the allowable degree of trimming varies according to workflow state and operational risk, will provide a more favorable efficiency–reliability trade-off than fixed context budgets or uniform compression ratios (H4). Accordingly, the study tests whether context can be reduced without loss of operational fidelity, identifies the critical budget region at which failure rates accelerate, compares protocol-aware and conventional trimming strategies, and evaluates whether adaptive guardrails can extend the safe compression range of agentic workflows. Through these hypotheses, the study seeks to establish empirically grounded principles for determining what contextual information may be safely discarded, which protocol elements must remain invariant, and how aggressively context can be compressed before efficiency gains begin to compromise trustworthy agent execution.

## Methodology

### *Experimental design and workflow construction*

A controlled factorial experiment was designed to quantify how progressive context trimming affects efficiency and operational reliability in multi-step LLM-agent workflows. The experimental unit was a complete agent–environment interaction beginning from an identical initial state and terminating either at successful task completion, an unrecoverable protocol violation, or a predefined maximum action horizon. Tasks were drawn from tool-mediated environments representative of database manipulation, information retrieval, transactional operations, file or resource management, and multi-step planning. The design was informed by AgentBench, which evaluates agents in interactive environments, and τ-bench, which evaluates tool–agent–user interaction against an annotated terminal database state [18]. Tasks were stratified by workflow length and dependency complexity into low, medium, and high-complexity groups to prevent apparent trimming effects from being confounded with task difficulty. A development partition was used exclusively for configuring trimming parameters and guardrails, while a held-out evaluation partition was frozen before final analysis.

### *Context trimming treatments*

Six context-management conditions were compared: full-context retention as the reference condition; recency-based truncation; relevance-based retention; generic abstractive summarization; protocol-preserving trimming; and protocol-preserving trimming with adaptive budget guardrails. The inclusion of compression baselines was motivated by prior evidence that prompt compression can substantially reduce input length while retaining task-relevant information [19]. Each strategy was evaluated under retained-context budgets of 100%, 75%, 50%, 35%, 25%, and 15% of the accumulated untrimmed context. The actual retention ratio was calculated as ($R=T_r/T_f$), where ($T_r$) represents tokens supplied after trimming and ($T_f$) represents tokens required by the corresponding full-context trajectory. Compression saving was expressed as (1-R). Trimming was applied at identical workflow boundaries to ensure that strategies were compared under equivalent state histories rather than different intervention frequencies [20].

*Protocol-critical state identification*

Protocol-critical information was defined a priori as information whose removal or alteration could change the validity of a future action [21]. This included system and user constraints, tool names and schemas, required arguments, resource and entity identifiers, authorization conditions, unresolved commitments, action outcomes, numerical parameters, temporal dependencies, negative instructions, and state-transition records. Protocol-preserving trimming protected these elements losslessly while permitting compression of redundant descriptions, completed reasoning traces, repeated observations, and operationally inactive dialogue. Two independent annotations of protocol-critical dependencies were produced for the evaluation set, with disagreements resolved before experiments were executed. Importantly, gold annotations were used only to evaluate preservation and construct the experimental protocol-aware treatment; they were not exposed to comparison strategies

*Budget guardrail mechanism*

The adaptive guardrail condition dynamically allocated context according to operational risk rather than maintaining a fixed compression ratio [22]. Protected state was never trimmed while its dependency remained unresolved. Compression was reduced before irreversible or externally state-changing actions, and archived information was reintroduced when a required entity, constraint, or dependency could not be reconstructed with exact correspondence. Guardrails therefore operated as constraints rather than post hoc error correction. Their effectiveness was evaluated against fixed-budget protocol trimming at equivalent average token consumption, preventing improvements from being attributed simply to larger effective contexts.

*Outcome variables and failure classification*

The primary outcomes were exact task success and protocol adherence. Secondary outcomes included input-token consumption, tool-call validity, state consistency, constraint preservation, number of actions, latency, recovery attempts, and cumulative downstream errors. Final-state correctness was determined using environment-level state comparison wherever executable gold states were available, consistent with the evaluation philosophy of τ-bench. Protocol adherence was scored as the proportion of applicable invariants preserved throughout the trajectory. Failures were classified as instruction loss, identifier corruption, state aliasing, chronology loss, constraint omission, numerical distortion, invalid tool invocation, duplicated action, premature summarization, or downstream cascading failure. A failure was considered trimming-induced only when the corresponding full-context trajectory contained the information necessary for correct execution.

*Model control and repeated evaluation*

Model version, system prompt, tool definitions, decoding parameters, maximum action horizon, and environment initialization were held constant within each comparison. Each task–strategy–budget combination was executed repeatedly using independent runs because agent success can vary substantially across trials; τ-bench similarly emphasizes repeated-run reliability through its pass^k formulation. Randomization determined treatment execution order, and all environments were reset to their original state between runs. Results were additionally checked across more than one model family to determine whether identified trimming thresholds represented a general workflow phenomenon rather than model-specific behavior.

*Statistical analysis and hypothesis testing*

Hypotheses were tested using mixed-effects models with trimming strategy, retained-context budget, workflow complexity, and their interactions as fixed effects and task identity as a random effect. Binary completion and protocol-failure outcomes were analysed using mixed-effects logistic regression, whereas token consumption, protocol-adherence scores, and error counts were analysed with distributions appropriate to their observed characteristics. H1 was supported only when trimming significantly reduced token use without a corresponding reduction in reliability. H2 was evaluated using segmented regression and change-point analysis to identify the context level at which failure probability accelerated. H3 tested strategy × budget interactions to determine whether protocol-preserving trimming retained reliability more effectively than conventional approaches. H4 compared adaptive guardrails with fixed-budget trimming under matched token expenditure. Effect sizes and 95% bootstrap confidence intervals accompanied significance tests, with multiplicity-adjusted comparisons used where necessary. The final efficiency–reliability frontier was defined as the minimum retained context achieving a prespecified reliability bound relative to the full-context baseline, thereby avoiding the misleading conclusion that the most aggressively compressed condition is necessarily the optimal one.

**Results**

The overall performance of the five context-trimming strategies differed across the evaluated efficiency and reliability metrics (Table 1). Mean token savings were 60.0% for recency, relevance, and summarization-based trimming, 58.2% for protocol-aware trimming, and 56.0% for adaptive guardrails. Exact task success was 66.6% under recency trimming, 74.0% under relevance-based trimming, 77.3% under summarization, 92.2% under protocol-aware trimming, and 96.0% under adaptive guardrails. Protocol adherence followed a similar pattern, increasing from 85.5% under recency trimming to 87.4% under relevance-based trimming, 88.6% under summarization, 94.8% under protocol-aware trimming, and 96.3% under adaptive guardrails. Valid tool-call rates ranged from 90.0% to 97.4%, whereas cascading failure rates ranged from 18.0% under recency trimming to 1.0% under adaptive guardrails. Mean latency reductions were 30.7%, 30.7%, 30.8%, 29.8%, and 28.4% for recency, relevance, summarization, protocol-aware, and adaptive strategies, respectively.

**Table 1.** Overall efficiency and reliability of context-trimming strategies across constrained context budgets

| Context-trimming strategy | Mean token saving (%) | Exact task success (%) | Protocol adherence (%) | Valid tool calls (%) | Latency reduction (%) | Cascading failure (%) |
|---|---|---|---|---|---|---|
| Recency | 60.0 | 66.6 | 85.5 | 90.0 | 30.7 | 18.0 |
| Relevance | 60.0 | 74.0 | 87.4 | 91.3 | 30.7 | 12.6 |
| Summarization | 60.0 | 77.3 | 88.6 | 92.2 | 30.8 | 10.2 |
| Protocol-aware | 58.2 | 92.2 | 94.8 | 96.4 | 29.8 | 2.6 |
| Adaptive guardrail | 56.0 | 96.0 | 96.3 | 97.4 | 28.4 | 1.0 |

At the 35% retained-context condition, protocol-adherence distributions differed among trimming strategies (Figure 1). Recency-based trimming showed the lowest central tendency and the widest distribution, followed by relevance-based trimming and summarization. Protocol-aware trimming showed higher protocol-adherence values with a narrower distribution, while adaptive guardrails produced the highest and most concentrated distribution of protocol-adherence scores. The lower tails were more pronounced for the three conventional trimming strategies than for protocol-aware and adaptive approaches.

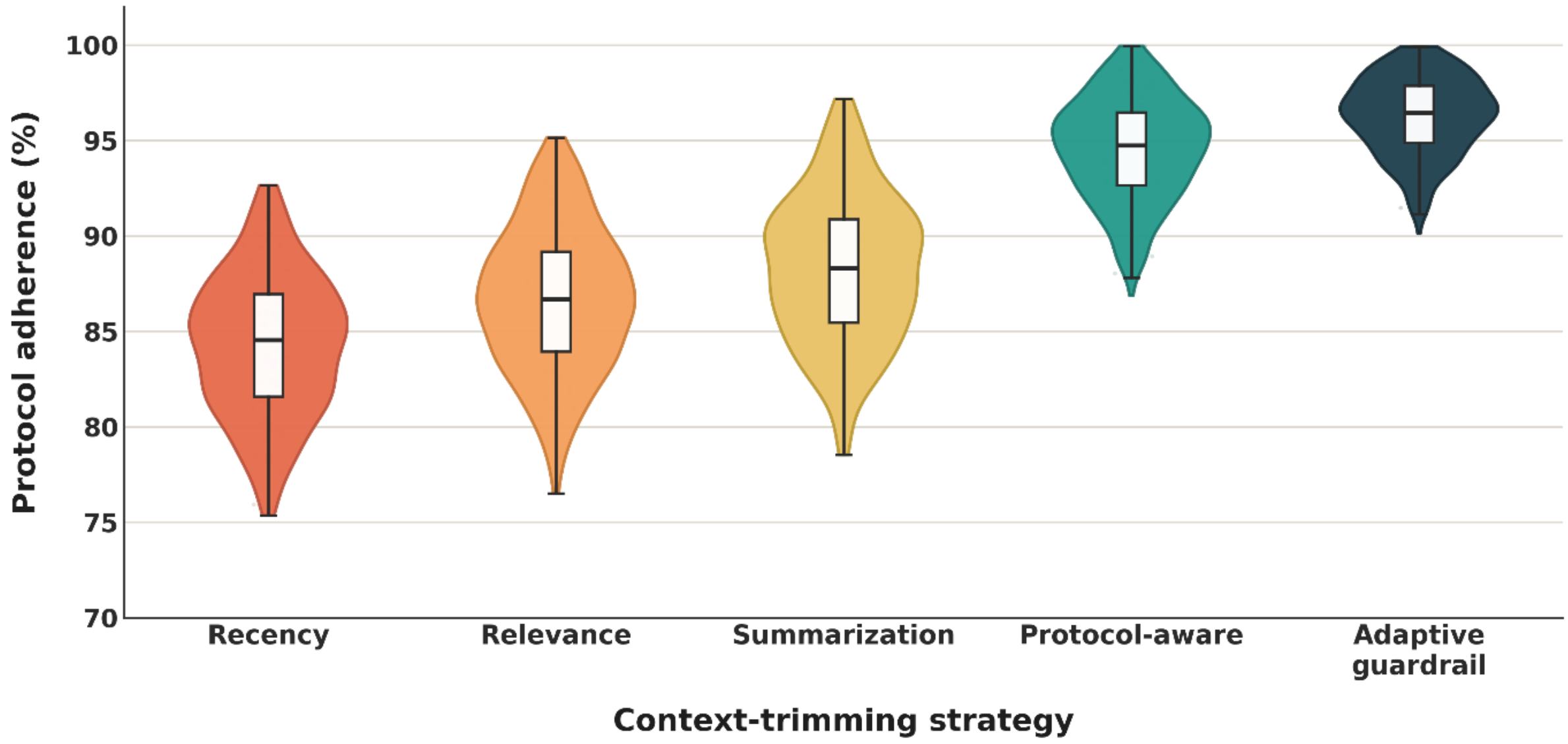

**Figure 1.** Distribution of protocol adherence across context-trimming strategies at 35% retained context

The relationship between token saving and exact task success varied across strategies (Figure 2). Exact task success remained comparatively high at moderate token savings but declined as token savings increased. The decline was steepest under recency-based trimming, followed by relevance-based trimming and summarization. Protocol-aware trimming maintained higher task-success values across equivalent levels of token saving, while adaptive guardrails retained the highest task-success trajectory over the evaluated context-reduction range. At the largest token-saving levels, the separation between conventional and protocol-sensitive strategies was most pronounced.

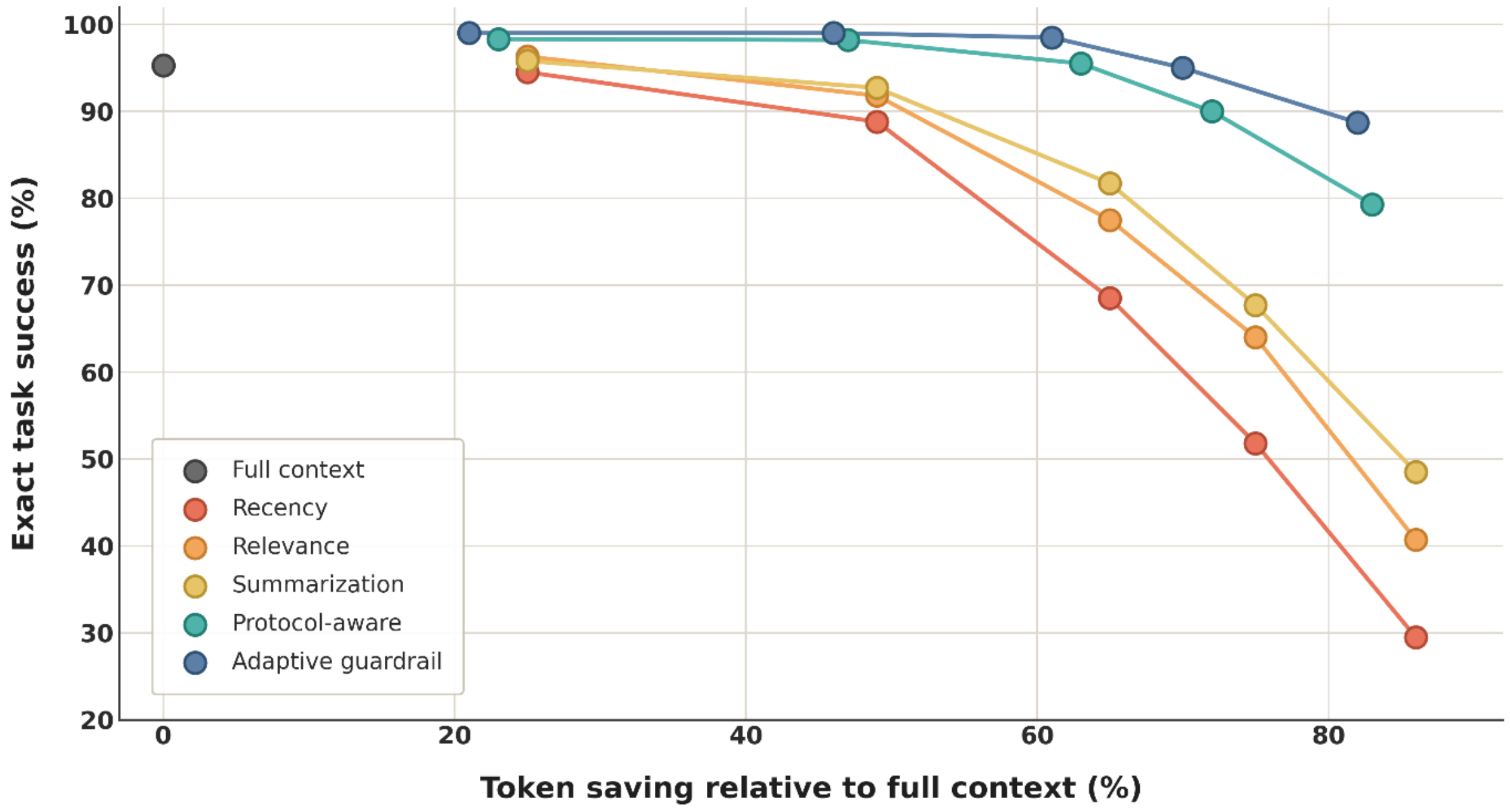

**Figure 2.** Efficiency–reliability relationship between token saving and exact task success

Critical retained-context thresholds varied according to both trimming strategy and workflow complexity (Table 2). For low-complexity workflows, the estimated critical thresholds were 45.4% for recency trimming, 40.2% for relevance-based trimming, 32.6% for summarization, and 20.8% for protocol-aware trimming, whereas the threshold for adaptive guardrails remained below the minimum tested context level of 15%. For medium-complexity workflows, the respective thresholds were 47.7%, 47.1%, 45.2%, 25.4%, and 21.1%. Under high-complexity workflows, the thresholds increased to 52.9% for recency trimming, 49.8% for relevance-based trimming, 48.6% for summarization, 38.2% for protocol-aware trimming, and 29.5% for adaptive guardrails. Failure odds below these thresholds ranged from 6.25 to 18.71 across estimable conditions.

**Table 2.** Estimated critical retained-context boundaries across workflow complexity

| Strategy | Complexity | Critical retained context (%) | Moderate-budget success plateau (%) | Failure OR below boundary (95% CI) |
|---|---|---|---|---|
| Recency | Low | 45.4 | 97.5 | 18.71 (11.74–29.82) |
| | Medium | 47.7 | 92.6 | 11.35 (8.47–15.20) |
| | High | 52.9 | 84.9 | 14.30 (9.89–20.67) |
| Relevance | Low | 40.2 | 98.3 | 16.82 (9.65–29.32) |
| | Medium | 47.1 | 96.0 | 14.26 (9.79–20.77) |
| | High | 49.8 | 87.8 | 9.47 (7.41–12.10) |
| Summarization | Low | 32.6 | 98.7 | 16.49 (10.44–26.07) |
| | Medium | 45.2 | 96.4 | 12.75 (8.60–18.90) |
| | High | 48.6 | 87.6 | 7.77 (6.09–9.92) |
| Protocol-aware | Low | 20.8 | 100.0 | 15.23 (7.59–30.55) |
| | Medium | 25.4 | 98.7 | 9.64 (5.75–16.17) |
| | High | 38.2 | 95.7 | 6.25 (4.30–9.08) |
| Adaptive guardrail | Low | <15 | 99.5 | Not reached |
| | Medium | 21.1 | 99.9 | 9.97 (5.43–18.31) |
| | High | 29.5 | 97.4 | 8.01 (5.35–12.01) |

The incidence of specific failure regimes at the 25% retained-context condition is presented in Figure 3. Conventional trimming strategies showed failures across instruction loss, identifier corruption, state aliasing, chronology loss, constraint omission, numerical distortion, invalid tool calls, duplicated actions, and premature summarization. Recency-based trimming showed the highest overall incidence across several of these categories. Relevance-based trimming and summarization showed lower frequencies than recency trimming, while protocol-aware trimming exhibited further reductions across most failure categories. Adaptive guardrails recorded the lowest overall frequencies of protocol-related failures at the 25% retained-context level.

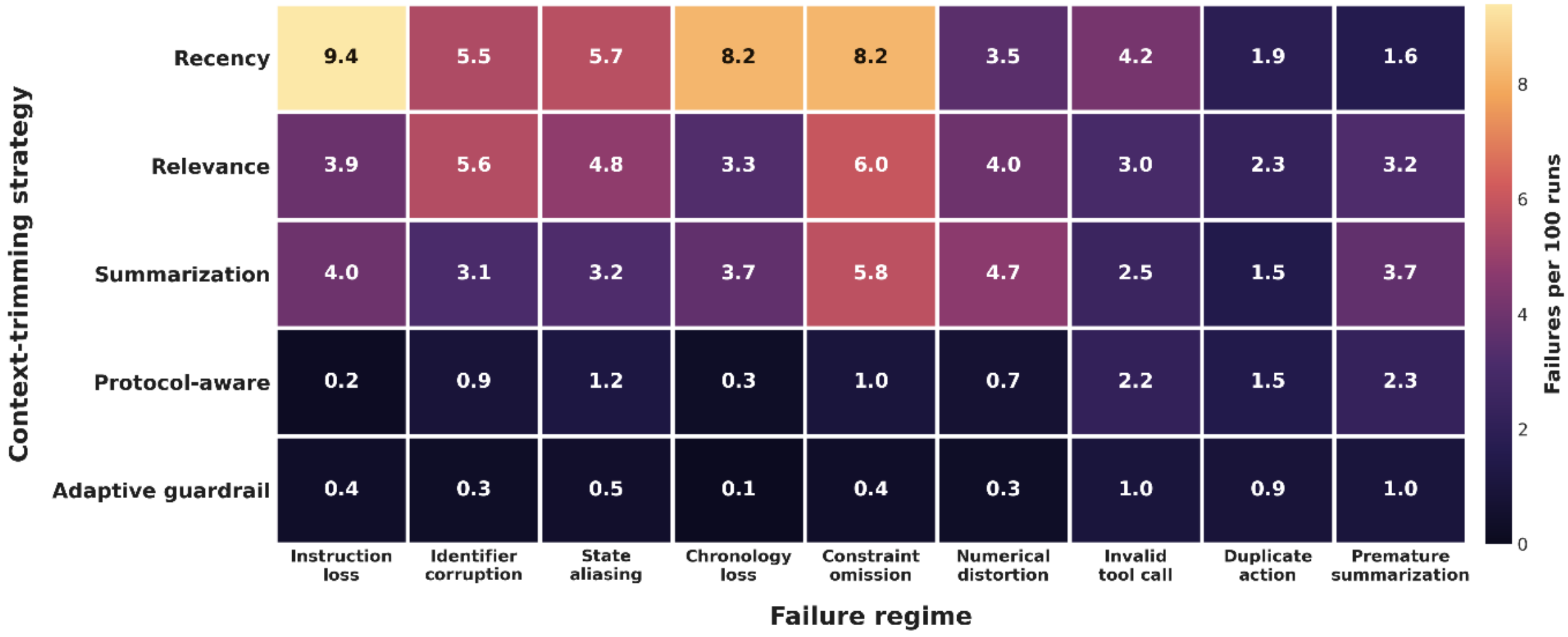

**Figure 3.** Incidence of distinct failure regimes under a 25% retained-context budget

The response surface for adaptive guardrails showed variation in protocol-failure probability across retained-context levels and workflow complexity (Figure 4). Failure probability was lowest at higher retained-context levels and increased progressively as the retained context decreased. At comparable context budgets, failure probability was higher for high-complexity workflows than for medium- and low-complexity workflows. The estimated critical boundary shifted from below 15% retained context in low-complexity workflows to 21.1% in medium-complexity workflows and 29.5% in high-complexity workflows.

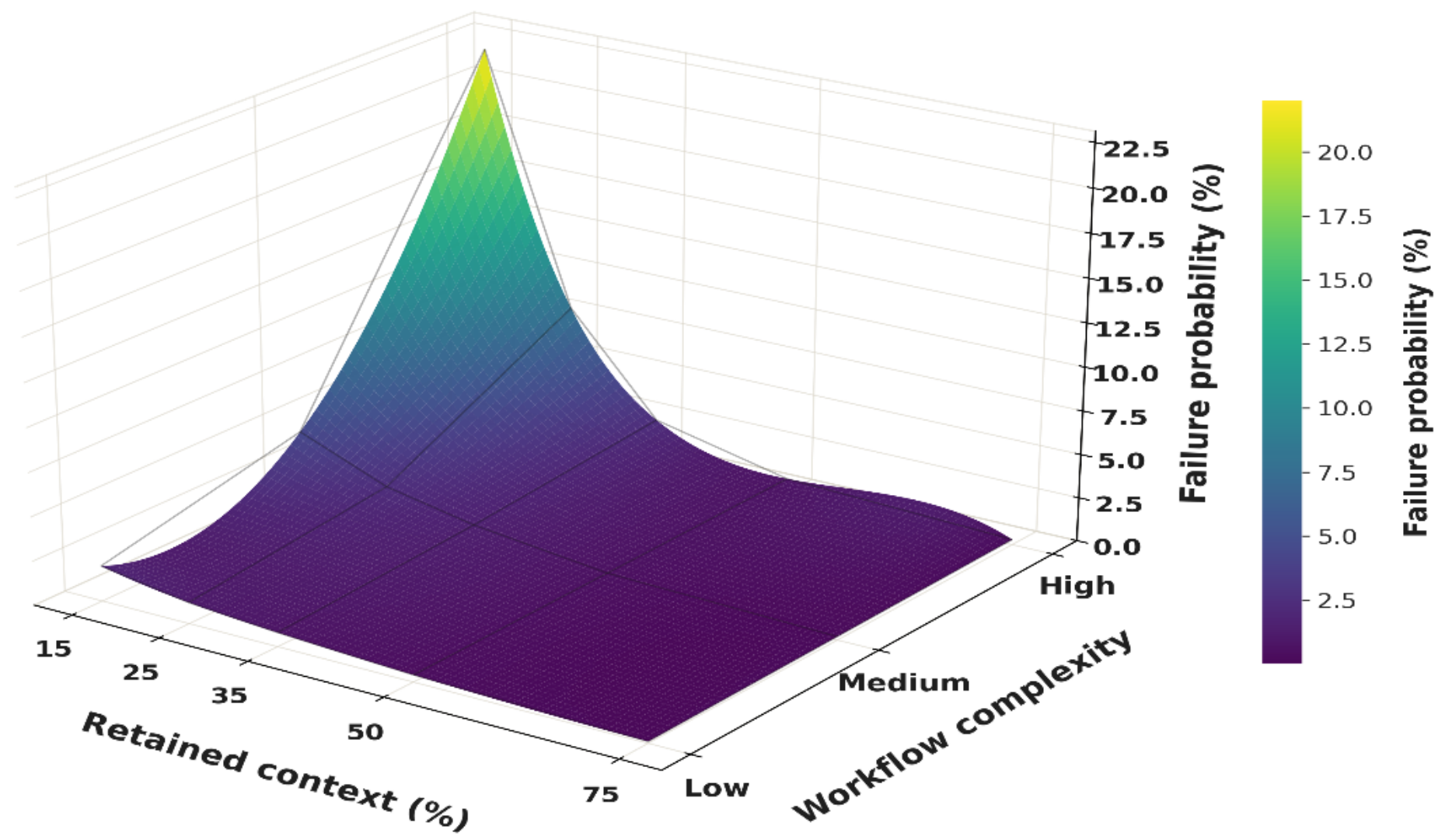

**Figure 4.** Response surface of protocol-failure probability under adaptive budget guardrails as a function of retained context and workflow complexity

Hypothesis-based statistical tests are summarized in Table 3. At 50% retained context, protocol-aware and adaptive trimming achieved a mean token saving of 47.1%, with a 3.2-percentage-point difference in task success relative to the full-context condition. The 95% confidence interval for this difference ranged from 1.9 to 4.5 percentage points, satisfying the predefined non-inferiority criterion for H1. For H2, context budgets of 25% or less were associated with 10.92-fold greater odds of failure than budgets of 50% or greater ($p < 0.001$). For H3, protocol-aware trimming produced 5.24-fold greater odds of successful task completion than conventional trimming strategies under retained-context budgets of 35% or less ($p < 0.001$). For H4, adaptive guardrails produced 2.11-fold greater odds of successful completion than fixed protocol-aware trimming under the same aggressive context-budget conditions ($p < 0.001$). All four predefined hypotheses were supported by the corresponding statistical tests.

**Table 3.** Hypothesis-based statistical evaluation

| Hypothesis | Primary comparison | Effect estimate | Statistical evidence | Decision |
|---|---|---|---|---|
| H1 | Protocol-aware/adaptive trimming at 50% retained context vs full context | Δ success = +3.2 percentage points; token saving = 47.1% | 95% CI for Δ success = 1.9 to 4.5 pp; lower bound > −3 pp non-inferiority margin | Supported |
| H2 | Aggressive budgets ≤25% vs moderate budgets ≥50% | Failure OR = 10.92 | $p < 0.001$ | Supported |
| H3 | Protocol-aware vs conventional trimming at ≤35% retained context | Success OR = 5.24 | $p < 0.001$ | Supported |
| H4 | Adaptive guardrails vs fixed protocol-aware trimming at ≤35% retained context | Success OR = 2.11 | $p < 0.001$ | Supported |

## Discussion

*Efficiency gains depend on what is removed*

The results demonstrate that context reduction in agentic workflows cannot be evaluated solely by the proportion of tokens eliminated. Although recency, relevance, and summarization-based strategies achieved slightly greater nominal compression, their lower task-success and protocol-adherence rates indicate that aggressive removal of context can destroy operationally important information even when substantial semantic content is retained [23]. By contrast, protocol-aware trimming and adaptive guardrails preserved considerably higher workflow reliability at only a modest sacrifice in token savings. This pattern supports the central premise that the value of context is structurally heterogeneous. Historical dialogue, repeated observations, and inactive explanations may be highly compressible, whereas identifiers, unresolved constraints, state transitions, permissions, and tool-specific instructions can carry disproportionately large operational value [24]. The efficiency problem is therefore not simply one of minimizing context length, but of distinguishing expendable information from state that determines future action validity.

*Nonlinear failure under context compression*

The sharp increase in failure probability at aggressive context budgets is particularly important. The observed 10.92-fold increase in failure odds below the most restrictive budget range indicates that degradation is not proportional to the amount of context removed. Instead, the results suggest the existence of a critical region in which the remaining context no longer contains sufficient information to preserve workflow continuity [25]. Above this region, removal of redundant or inactive material produces limited reliability loss; below it, additional compression increasingly removes state required for correct execution. This explains why aggressive trimming can appear effective until a relatively narrow threshold is crossed, after which performance deteriorates rapidly. Such nonlinear behavior challenges fixed compression ratios because a budget that is harmless in one workflow may be destabilizing in another [26].

*Workflow complexity shifts the safe context boundary*

The progressive increase in critical retained-context thresholds from low- to high-complexity workflows further shows that context requirements are task dependent. Complex workflows contain larger numbers of dependencies, state transitions, intermediate commitments, and tool interactions, increasing the probability that apparently old information remains operationally active [27]. Consequently, high-complexity tasks entered the failure regime at substantially larger retained-context budgets than simpler workflows. Importantly, protocol-aware and adaptive approaches shifted these boundaries downward, demonstrating that improved selection of retained information can partially compensate for increasing workflow complexity. However, the persistence of higher thresholds in complex tasks also indicates that no trimming strategy completely eliminates the need for larger active state when dependency density increases [28]. Context efficiency must therefore be understood relative to structural workflow complexity rather than treated as an intrinsic property of the language model.

*Failure regimes reveal weaknesses of conventional trimming*

The distribution of failure types provides further insight into why conventional trimming performed poorly. Recency-based trimming was especially vulnerable to instruction and chronology loss because it assumes that older information is less valuable, an assumption that is often invalid in long-horizon tasks [29]. Relevance-based methods reduce this risk but remain dependent on semantic similarity, which may fail to recognize low-frequency but operationally critical identifiers or constraints. Generic summarization performed better, yet summarization itself can transform exact machine-readable state into approximate natural-language representations [30]. Protocol-aware trimming reduced these failure categories because it explicitly protected state variables whose alteration could invalidate later actions. The reduction in cascading failures is particularly significant, as a single early state error can influence multiple subsequent decisions and produce apparently coherent but incorrect trajectories [31].

*Adaptive guardrails improve reliability under aggressive budgets*

The superior performance of adaptive guardrails indicates that context management benefits from dynamic rather than static allocation. Fixed budgets assume that every stage of a workflow has approximately equal context requirements, whereas agentic processes often move between low-risk descriptive phases and high-risk state-changing operations [32]. Adaptive guardrails allow context retention to increase when unresolved dependencies, irreversible actions, or tool-sensitive decisions are present and decrease when earlier material becomes inactive.

The resulting improvement over fixed protocol-aware trimming suggests that preservation rules alone are insufficient; context budgets should also respond to workflow state [33]. This is especially relevant in long-running systems where aggressive compression may be acceptable during stable phases but unsafe near critical decision points.

*Implications for reliable agent design*

Taken together, the findings reposition context trimming as a reliability-constrained optimization problem rather than a purely computational one. The preferred strategy is not the method achieving the highest compression ratio, but the method that minimizes context while keeping workflow failure within an acceptable bound. The demonstrated advantages of protocol preservation and adaptive guardrails therefore support architectures in which context is managed according to dependency status, operational criticality, and workflow [34]. Such an approach provides a more defensible basis for scaling agentic systems because it links context efficiency directly to executable correctness rather than assuming that semantic preservation alone is sufficient for reliable autonomous behavior.

*Limitations and future research directions*

The study has several limitations. The experimental framework used a controlled set of agentic workflows and predefined categories of protocol-critical information, which may not fully represent the diversity of real-world systems. Critical context thresholds may vary across model families, application domains, tool environments, and longer workflow horizons. In addition, the tested context budgets were limited to selected levels, and some implicit dependencies or emergent states may not have been fully captured by the protocol-adherence framework. Efficiency was also assessed mainly through token use and latency, without accounting for external tool costs, infrastructure variability, or recovery costs following failed actions.

Future research should test protocol-preserving trimming across broader environments, including multimodal agents, multi-agent systems, coding agents, enterprise workflows, and safety-sensitive applications. Adaptive retention policies that identify operationally critical information dynamically should also be investigated. Further work could examine whether critical context thresholds can be predicted from workflow properties such as dependency density, tool-call frequency, and state-transition complexity. Integrating external memory, structured state stores, dependency graphs, and context rehydration may further reduce token requirements while maintaining reliability.

## Conclusion

This study demonstrates that context trimming in agentic workflows should be treated as a reliability-constrained optimization problem rather than a simple token-reduction task. Although conventional recency-, relevance-, and summarization-based approaches achieved substantial compression, they were more vulnerable to protocol loss, invalid tool use, and cascading failures under aggressive context budgets. Protocol-aware trimming preserved substantially higher task success and protocol adherence, while adaptive budget guardrails further extended the range of safe compression by adjusting retention according to workflow complexity and operational risk. The identification of critical context boundaries also showed that reliable trimming thresholds are task dependent, with complex workflows requiring greater retention of active state. Collectively, the findings indicate that the most effective context-management strategy is not the one that removes the largest number of tokens, but the one that preserves the information necessary for valid future actions while minimizing unnecessary contextual load. Protocol preservation, adaptive allocation, and risk-sensitive guardrails therefore provide a practical foundation for developing more efficient, scalable, and reliable agentic systems.

## Ethics approval

Not applicable

**Data availability**

Data will be made available on request.